\documentclass[10pt,letterpaper]{article}

\usepackage[english]{babel}
\usepackage{cogsci}
\usepackage{apacite}
\usepackage[pdftex]{color,graphicx}
\usepackage{amsmath, amsthm, amssymb}
\usepackage{rotating}
\usepackage{caption}
\usepackage{subcaption}
\usepackage{subfig}

\title{Evolving useful delusions: Subjectively rational selfishness leads to objectively irrational cooperation}

\author{{\large \bf Artem Kaznatcheev (artem.kaznatcheev@mail.mcgill.ca)} 
  \\ School of Computer Science and Department of Psychology, McGill University, 
  \\ 3480 University Street, Montreal, QC H3A 0E9 Canada \\
  \AND {\large \bf Marcel Montrey (marcel.montrey@mail.mcgill.ca)}
\\ Department of Psychology, McGill University, 
\\ 1205 Penfield Avenue Montreal, QC H3A 1B1 Canada
 \AND {\large \bf Thomas R. Shultz (thomas.shultz@mcgill.ca)}
\\ Department of Psychology and School of Computer Science, McGill University,
\\ 1205 Penfield Avenue Montreal, QC H3A 1B1 Canada
}

\begin{document}

\maketitle

\begin{abstract}
We introduce a framework within evolutionary game theory for studying the distinction between objective and subjective rationality and apply it to the evolution of cooperation on 3-regular random graphs. 
In our simulations, agents evolve misrepresentations of objective reality that help them cooperate and maintain higher social welfare in the Prisoner's dilemma. 
These agents act rationally on their subjective representations of the world, but irrationally from the perspective of an external observer.
We model misrepresentations as subjective perceptions of payoffs and quasi-magical thinking as an inferential bias, finding that the former is more conducive to cooperation.
This highlights the importance of internal representations, not just observed behavior, in evolutionary thought.
Our results provide support for the interface theory of perception and suggest that the individual's interface can serve not only the individual's aims, but also society as a whole, offering insight into social phenomena such as religion.

\end{abstract}

\section{Introduction}

Economic theory has traditionally assumed it impossible, or at least unreasonably difficult, to directly observe preferences and beliefs, preferring to assess agents by their behavior.
A similar tradition exists in evolutionary biology and ecology, where internal factors are seen as secondary to the behavior that determines fitness and thus evolutionary outcomes.
Given these roots, it is unsurprising that evolutionary game theory (EGT) has adopted a similar attitude, embracing a behavior-centric approach to describing model agents.
While this has made a great deal of theoretical work tractable, it has also made human deviations from rationality difficult to model.
Our primary goal is to overcome this limitation by introducing into EGT a framework for studying differences between objective and subjective experiences.

Of central importance to classical economic theory is the simplifying assumption of rationality. 
In traditional theory, this rationality is taken literally and is assumed to apply at the objective level of explicit payoffs and behavior.
However, if we are to model humans, we should be conscious of the fact that decisions act on internal representations of rewards and beliefs about the consequences of past and future actions.
There is no \emph{a priori} reason to believe that these internal representations are perfectly aligned with external payoffs.
In fact, the experimental field of neuroeconomics reveals that there are often marked differences between the objective and subjective rewards that participants experience~\cite{L08}. 
Decisions made on the basis of these internal representations can thus appear irrational to an external observer, who is na\"{i}ve about the agent's mental state and only aware of the external rewards.
However, from the agent's perspective, these actions could be rational, given the agent's internal representation. This is the concept of subjective rationality.
It is reasonable to explore the possibility that many of these subjective representations are shaped by evolution.

To illustrate this concept, we look at  cooperation---the typical proving ground for EGT. 
Because cooperation involves paying a cost so that another individual can derive a benefit, cooperators risk being exploited.
Often, as in the Prisoner's dilemma (PD) game, the objectively rational strategy is to defect rather than cooperate, because defection exploits cooperators and prevents exploitation at the hands of other defectors.
Nevertheless, humans playing PD often cooperate, even when experimenters are careful to exclude factors that encourage cooperation~\cite{S95,FF03}.
Research suggests that this is caused by a propensity to consider extraneous subjective factors, such as uncertainty~\cite{ST92,C99}, 
irrelevant goals~\cite{BM06,SLL08}, aversion to inequality~\cite{FS99,FC07}, discounting of future rewards~\cite{ER98,LCMB04}, counterfactual reasoning~\cite{LMCM07,C11}, in-group favoratism~\cite{K10b,HKS13}, and judgments of reputation~\cite{NS98}, trustworthiness~\cite{MSDH09} and other moral characteristics~\cite{DFP05}.
We argue that, while such behavior is objectively irrational, it could nevertheless be rational from a subjective point of view.
Human participants may be maximizing their payoffs; it's just that the game they are perceiving may not be the one that the experimenter intended.
By directly modeling the agent's subjective state, learning and decision making process, we are embracing McNamara's~\citeyear{M13} suggestion that researchers should consider  enriching EGT models with features such as psychological mechanisms, decision making, personality variations and novel traits, with the aim of accounting for the extensive variation in empirical studies across both people and cultures~\cite{FF03,HBBCFG04}.

\subsection{Quasi-magical thinking}

An interesting category of subjective effects is what Shafir and Tversky~\citeyear{ST92} termed quasi-magical thinking (QMT).
Whereas magical thinking refers to the mistaken belief that an action affects an outcome that it cannot, QMT describes situations where behavior is consistent with this belief without it being explicitly held.
Several counterintuitive findings in behavioral game theory fall under this heading.
For instance, humans cooperate more readily when a partner's choice is unknown~\cite{ST92}---a puzzling response pattern consistent with the erroneous view that cooperation will encourage reciprocity, even in one-shot games.
Similarly, participants make more optimistic predictions about their partner's chances of cooperation if they themselves have already decided to cooperate~\cite{DMS77}, as if their decision can influence their partner.

Though QMT seemingly violates rationality, Masel~\citeyear{M07} argues that it could result from a reasonable inferential bias.
When judging a partner's probability to cooperate, assuming that that partner is similar to oneself can yield a rapid, accurate prediction if information is scarce.
A player might therefore benefit from observing not just what a partner does, but also what they would have done in the partner's position.
If the player is indeed similar to the population at large, then this bias could yield valuable data.
Masel~\citeyear{M07} showed that, if used during periods of uncertainty, this type of QMT could also lead to greater cooperation in a public-goods game.

In the next section, we present a novel EGT framework for representing and acting on subjective conceptions of reality. 
We use this framework to demonstrate that agents playing PD on a random 3-regular graph, when allowed to evolve subjective representations of the game, converge on an objectively incorrect representation of the interaction.
This misrepresentation of the objective game's payoffs leads our subjectively rational agents to cooperate---an objectively irrational choice.
We further show that an inferential bias (QMT) also promotes cooperation, though to a lesser extent than misperceiving the payoffs does.
This is our contribution to the mounting evidence that rationality and irrationality should not be dichotomized and that internal representations are important to evolutionary dynamics.
While humans certainly behave irrationally in PD from an objective perspective, in that they do not optimize the stated payoffs, they might be using a perceptual interface where subjective payoffs reflect extraneous concerns in addition to individual fitness effects.
They might also be using QMT as a reasonable inferential bias.
Humans, in other words, may be acting rationally on their subjective rather than objective payoffs or using subjective knowledge in unanticipated ways.


\begin{figure}
	\center
    \includegraphics[width=\columnwidth]{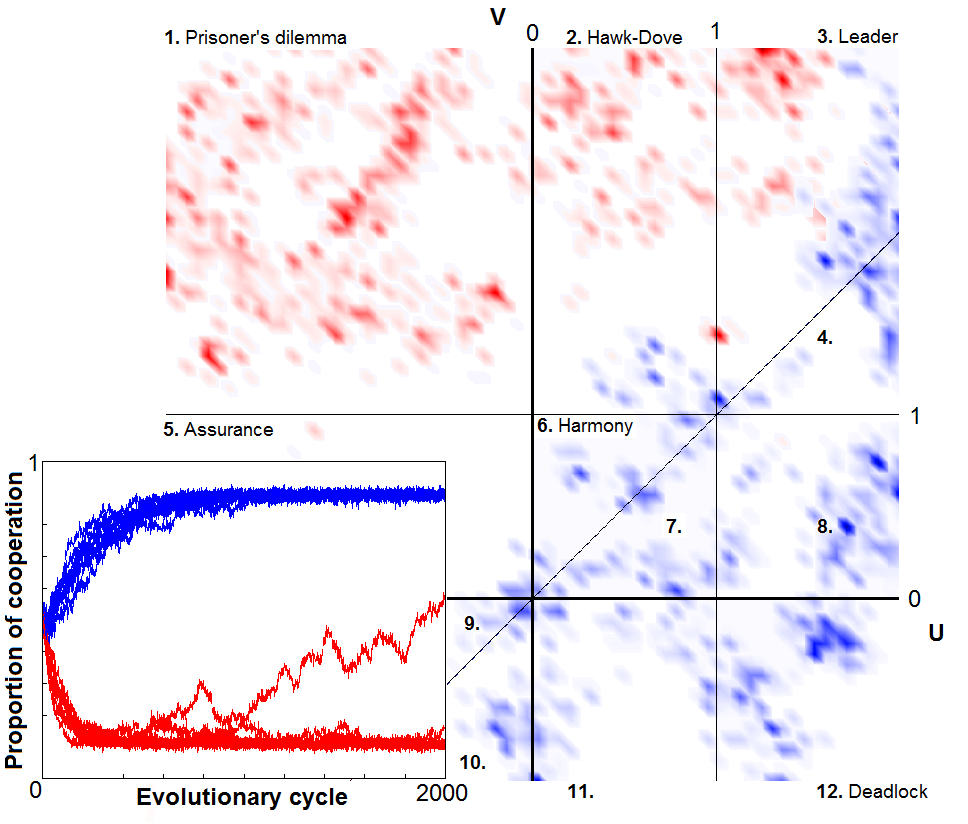}
    \caption{U-V plane partitioned into 12 regions according to rank ordering of the payoffs; notable games are named. 
    Shading corresponds to the density of agents' evolved subjective payoffs, with the blue agents from runs with objective reality of $(U,V) = (-0.3,1.3)$ and red from $(U,V) = (-0.9,1.9)$. 
    A total of 10000 agents from 20 independent runs are plotted at the 2000th time step. 
    The inset shows the proportion of cooperation versus evolutionary cycle, where each line is a run.}
    \label{fig:densePlot}
\end{figure}

\section{Model and Method}
\subsection{Game space}

In our model, an agent (Alice) engages in a game with an adjacent individual (Bob).
The players independently choose between two pure strategies: to cooperate (C) or defect (D).
By subtracting from all payoffs any constant offset and choosing our units, the payoff for Alice is given by the canonical matrix $\begin{pmatrix}1 & U \\ V & 0 \end{pmatrix}$ where Alice's choice determines the row (first row for C, second for D) and Bob's the column (first column for C, second for D). 
We can plot this parameterization of all two-player symmetric games in the U-V plane, as shown in Figure~\ref{fig:densePlot}. 
The different possible rank orderings of the $4$ payoffs divide the plane into $12$ qualitatively distinct regions. 
Some of the well-known games are marked with their names; game 1, for instance, is Prisoner's dilemma (PD) and has $V > 1 > 0 > U$. 
The typical representation of the PD game, in terms of a cooperator providing a benefit ($b$) to her partner at a cost ($c$) to herself, with $b > c > 0$, translates to $(U,V) = (-\frac{c}{b - c}, \frac{b}{b - c})$ in the canonical representation. 
A not-so-competitive (i.e., friendly) environment has $\frac{c}{b - c}$ near 0, and a competitive envrionment has $\frac{c}{b - c} \geq 1/2$. 
The exact value of the right hand side depends on factors like the spatial structure and can be calculated from Ohtsuki-Nowak transform~\cite{ON06}; $1/2$ corresponding to the 3-regular random graph condition that we study.

If $V < 1$ (games 5--12), then cooperation is rational (and an evolutionarily stable strategy (ESS)), and if $U < 0$ (games 1,5,9 and 10) then defection is rational (and ESS); otherwise both pure strategies are irrational without more information about the partner. 
For games 5 (Assurance), 9 and 10 (i.e., when both $U < 0$ and $V < 1$), cooperation and defection are both rational, with the rational behavior depending on further information about the partner.

\subsection{Subjective representations}

For each run of the simulation, we fix a specific game $(U,V)$ as objective reality.
This game's payoffs determine agents' fitness and thus drive evolution.
However, for the agent's decision process (cooperation vs. defection), we do not allow explicit access to these objective payoffs.
Instead, each agent has an evolved internal representation of the game---their individual perception of the payoffs.
For instance, Alice might think the game is $(U_A,V_A)$ and Bob might think it is $(U_B,V_B)$.
Suppose Alice cooperates and Bob defects.
Their fitness is adjusted according to the real game, so Alice's fitness changes by $U$ and Bob's by $V$.
However, Alice believes that the effect on her fitness is $U_A$ and Bob thinks the effect on his is $V_B$.

These internal conceptions of the game can be misperceptions, emotional biases like inequality aversion, or other ingrained beliefs about the underlying interaction. 
The representation of the game passes from parent to offspring under the influence of natural selection. 
Successful agents pass on their conception of the game (given by a value $-2 < U < 2$ and $-1 < V < 3$), subject to a small mutation rate: With probability $0.05$ (mutation rate), the generational transmission is faulty and the offspring is born with a game selected uniformly  at random from $\pm 0.1$ (mutation size) the parent's $U$ and $V$ values. 
The qualitative results are robust to changes in mutation, as long as the rate is not unreasonably high.

\subsection{Learning and quasi-magical thinking}

The last ingredient for decision making is estimating the probability that a partner will defect or cooperate. 
Alice has a mind that consists of her estimate $\hat{p}_A$ of the probability $p$ that her partner cooperates when she cooperates, and $\hat{q}_A$ of the probability $q$ that her partner cooperates when she defects.
Since $(U_A, V_A)$ do not change during Alice's lifetime, we count it as genetically determined and shaping the mind, but not a direct part of it.
No agent can condition their action on the action of their partner, $p = q$, but we avoid hard-coding this knowledge into Alice and allow $\hat{p}_A$ to differ from $\hat{q}_A$.

The agent is thus fully specified by 4 parameters: her perceived game $(U_A, V_A)$
and her probability estimates $\hat{p}_A$ and $\hat{q}_A$. 
She acts rationally on this information, deciding to cooperate only if the expected subjective utility of doing so is greater than that of defection, $\hat{p}_A + (1 - \hat{p}_A) U_A > \hat{q}_A V_A$.

To ensure that Alice has a chance to sample both strategies, we incorporate a small trembling-hand parameter~\cite{S75}:
With probability $\epsilon = 0.1$, Alice takes the action opposite of the one she intends.

Alice's estimates of $\hat{p}_A$ and $\hat{q}_A$ are based on Bayesian inference from an initially uniform distribution, resulting in rational learning, or calculation of the maximum likelihood estimate~\cite<also the best Bayesian estimate given the squared error loss function;>{GY06}:

\begin{equation}
\hat{p}_A = \frac{n_{CC} + 1}{n_{CC} + n_{CD} + 2}, \;
\hat{q}_A = \frac{n_{DC} + 1}{n_{DC} + n_{DD} + 2}
\end{equation}

For QMT agents, we adopt a strategy similar to \citeA{M07} and modify the mechanism for belief updating. If Alice is a QMT agent, then she not only observes what Bob does, but also simulates what she would have done in his place and then uses both pieces of information in her Bayesian learning. This results in:

\begin{equation}
\hat{p}_A = \frac{2n_{CC} + n_{CD} + 1}{2(n_{CC} + n_{CD}) + 2}, \;
\hat{q}_A = \frac{n_{DC} + 1}{2(n_{DC} + n_{DD}) + 2}
\end{equation}



In both sets of equations, $n_{ij}$ is the number of times Alice acted $i$ and her partner acted $j$.
The agents are simple Bayesian learners that update their minds after each interaction. A newborn agent has $n_{CC} = n_{CD} = n_{DC} = n_{DD} = 0$ and thus $\hat{p} = \hat{q} = 1/2$.

\subsection{Structured interactions}

Agents and the strategies they implement do not work in a vacuum: An agent's payoff is a function of both its strategy and the context in which that strategy is executed. 
The spatial (or interaction) structure of the world is thus central to the question of cooperation and altruism in EGT~\cite{AB02,SF07}. 
In fact, without any interaction structure (an inviscid environment), no matter what reasonable representation of the agents' behavior we choose, cooperation will not emerge in PD. 

We consider a minimal spatial structure and generate random $3$-regular graphs on 500 nodes.
Using the analytic theory of Ohtsuki and Nowak~\citeyear{ON06} for analyzing these graphs, we expect to find cooperation when the inverse of the competitiveness is between $0$ and $1/2$. 
When $\frac{c}{b -c} = \frac{1}{2}$, we expect a rapid phase transition from universal cooperation to universal defection.

Agents inhabit the nodes of the graph (one agent per node) and interact with adjacent agents.
The simulation begins from a random distribution of agents over the graph and over genetic space ($-2 < U < 2$, $-1 < V < 3$ and whether the agent uses regular Bayesian inference or QMT).
Each evolutionary cycle alternates between generating fitness from interactions and reproducing. 
During the interaction step, Alice decides her action independently for each neighbor, updating her mind ($\hat{p}_A$ and $\hat{q}_A$) after each interaction. 
At each reproductive step, $10\%$ of agents are randomly selected for death.
This death rate acts as a seperation of the interaction and evolution time-scales; with $10\%$ and $3$-regular graphs, an average agents has about $30$ interactions during its life, but the results are largely unchanged for reasonable (not too high) death rates.
Neighbors of the perished agents compete to repopulate the vacated cells with their offspring~\cite<known as death-birth updating;>{ON06}. 
The probability of Alice winning this competition is proportional to the objective payoff she accumulates from the current round of interactions with all her neighbors.

\begin{figure}
	\center
	\includegraphics[width=\columnwidth]{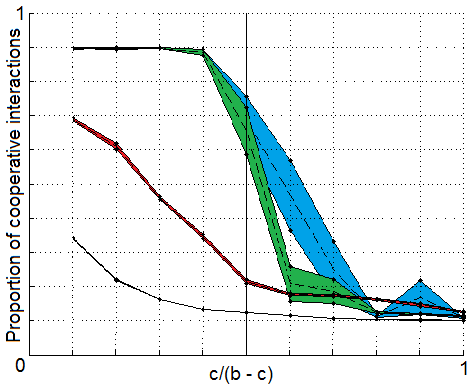}
	\caption{Proportion of cooperative interactions versus $\frac{c}{b - c}$. Blue shows the evolution of subjective payoffs; red, evolution of standard Bayesian and QMT agents; green, co-evolution of both; and black, completely rational Bayesian agents that do not evolve. Line thickness represents standard error from averaging 10 runs.}
	\label{fig:conditions}
\end{figure}


\section{Results}
 
Our primary results are presented in Figure~\ref{fig:densePlot}. 
The main figure is a density plot of agents by their genetic internal representation of the game. 
Darker regions correspond to more agents with those $(U,V)$ values. 
A total of 5000 agents of each color are plotted, recorded from the last evolutionary cycle of 10 independent simulations. 
The objective game is PD, with $(U,V) = (-0.3,1.3)$ in blue, and $(U,V) = (-0.9,1.9)$ in red.

 
For the highly competitive world indexed by red data points, where the benefit of cooperation is only $19/9$ times the cost, most agents evolve toward an objectively correct conception of the game: The internal representation of most of the agents corresponds to PD, with a minority evolving toward Hawk-Dove and Leader games.
On the other hand, in the relatively friendly environment indexed by blue data points, where the benefit of cooperation is only $13/3$ times the cost, most agents evolve toward misrepresentations of objective reality that are conducive to cooperation. 
In particular, most agents in the blue condition evolve internal representations where the only rational action is cooperation: Games 6-8 and 11-12. 
No agents in the friendly environment evolve the objectively correct PD representation.


The benefit of misrepresenting reality is highlighted in the inset of Figure~\ref{fig:densePlot}, which shows the proportion of cooperation versus evolutionary cycle. 
Blue lines correspond to the proportion of cooperation in 10 independent runs with $(U,V) = (-0.3, 1.3)$ and the red lines correspond to 10 runs with $(U,V) = (-0.9, 1.9)$. 
Although the actions of agents in the friendly environment are objectively irrational, they are subjectively rational and---through their subjective misrepresentation of reality---produce much higher levels of cooperation and greater social welfare than do the agents in the highly competitive environment.

Introducing QMT highlights the relative importance of perception vs. inference.
As can be seen in Figure~\ref{fig:conditions}, the co-evolution of perception (subjective payoffs) and inference (standard vs. QMT) results in the green curve that best approximates the expected transition from all cooperation to all defection (offset of $0.1$ from total saturation due to shaky hand in both cases) at $\frac{c}{b - c} = \frac{1}{2}$.
However, subjective payoffs and QMT evolving on their own relax the transition in different directions. 
In particular, if perception is fixed to objective truth and inference is allowed to evolve then we get significationly lower levels of cooperation in the low competition regime (red line). 
On the other hand, if perception evolves but inference is fixed at rational (no possibility for QMT) then cooperation is sustained in more competitive regimes than expected. 
The social welfare from misperception is greather than that of QMT, although both lead to significant cooperation in friendly environments.

\section{Discussion}

Our results demonstrate several key points concerning the relationship between behavior and internal representations. 
To start, it is not necessarily the case that agents evolve a true representation of the game they are playing.
This agrees with findings in behavioral game theory suggesting that humans often deviate from objectively rational behavior by incorporating seemingly extraneous factors into their decision-making process~\cite{FC07,SLL08,L08}.
Use of such information is analogous to misinterpreting the game's payoff matrix---something that our simulated agents do when the world is less competitive.

However, we also observe that misrepresentation is not universal. 
In highly competitive environments, agents evolve an approximately correct representation of objective reality and do not succumb to QMT. 
This suggests an empirical test of our theory: Do humans misrepresent reality more in friendly social settings or competitive ones? 
Our study suggests that mental representations are more accurate in the latter case. 

Furthermore, we show that misrepresenting reality is not necessarily detrimental to individual or group performance. 
In the friendly environment, agents' misrepresentations allow cooperation to flourish, despite each agent being solely interested in maximizing its own payoff.
This allows the population to overcome cooperation's risk of exploitation because the cooperators become insensitive to the possibility of defectors taking advantage of them.
Inevitably, mutual cooperation emerges and, in the friendly environment, is sufficient to sustain erroneous perception of payoffs.

\subsection{Realist vs. interface theories of perception}

The most direct consequence of our results is for understanding the veridicity of perception. 
The orthodox view~\cite{YB96,P99} is \emph{critical realism}---perception resembles reality, although it does not capture all of it. 
The typical evolutionary justification for this is that veridicity has greater utility for an agent and will be selected for by natural selection. 
Our results show that this is not always the case.

Hoffman~\citeyear{H98,H09} provides an alternative, by hypothesizing that perception is an interface that hides unnecessary complexity irrelevant to the agents' aims.
In the case of evolution, the ``aim'' is maximizing fitness, and thus perception does not need to be truthful, but has to provide an interface through which the agent can act to maximize its fitness. 
Mark et al.~\citeyear{MMH10} confirmed this with an evolutionary model showing that fitness is more important than ``truth'' to the agent. 
If perception is expensive, then the agent will tune it to reflect the fitness distribution---something that depends on the interaction between agent and objective reality.

Our results extend beyond this to show that sometimes tuning reflects not just the individual agent's, but the population's interaction with the world. 
Our agents evolve misrepresentations of objective reality that provide them with incorrect fitness information to promote a social good. 
This happens despite the fact that the agents act rationally on their perceptions. 
Further, unlike Mark et al.~\citeyear{MMH10}, we don't have to include a penalty for more accurate representations. 
In fact, our agents are capable of overcoming an implicit penalty associated with misrepresentation. 
In the less competitive environments where cooperation emerges, if Alice were to suddenly switch to accurate perception of payoffs then---in the short term---she could exploit her neighbors to get a strictly higher fitness.
In short, our results not only strengthen the case for the interface theory of perception, but also suggest regarding the individual's interface as not just serving the aims of that individual, but those of society as a whole.

An interesting domain for such social interfaces is religion. 
Religion is often lauded for promoting cooperation and moral behavior~\cite{B03,RB06} 
and often criticized for disseminating incorrect and even delusional beliefs~\cite{D06,H07}.
When evaluating the net impact of religion, these two well-supported positions are typically placed in opposition. 
Our model is consistent with both of these claims, while providing an explanation of how these tendencies can emerge from the same underlying process.  
Unlike previous work~\cite{RR03,JB06}, our model does not rely on group-selection or punishment, so it applies to both moralizing and non-moralizing gods, reaching more cultures.

To sum up, by creating agents who lack an \emph{a priori} understanding of the world, we demonstrate that such agents can evolve a misrepresentation of reality. 
Furthermore, because evolution selects for adaptive behavior rather than accurate internal representations, these delusions may prove useful by encouraging a greater degree of cooperation than rationality would otherwise allow. 
By offering an example of how internal representations and their consequences for behavior can be studied in a game theoretic context, we hope to pave a path for understanding how and why humans deviate from objective measures of rationality. 
Though often ignored, the key may be subjective experiences.

\noindent{\tiny {\bf Acknowledgements:} This work is supported by a grant from the Social Sciences and Humanities Research Council of Canada to TRS.}

\bibliographystyle{apacite}

\setlength{\bibleftmargin}{.125in}
\setlength{\bibindent}{-\bibleftmargin}
\bibliography{references}

\end{document}